# Developing an Interactive Tutorial on a Mach-Zehnder Interferometer with Single Photons

Chandralekha Singh and Emily Marshman

*Department of Physics and Astronomy, University of Pittsburgh, Pittsburgh, PA, 15260, USA*

**Abstract:** We are developing a Quantum Interactive Learning Tutorial (QuILT) on a Mach-Zehnder Interferometer with single photons to expose upper-level students in quantum mechanics courses to contemporary applications. The QuILT strives to help students develop the ability to apply fundamental quantum principles to physical situations and explore differences between classical and quantum ideas. The QuILT adapts visualization tools to help students build physical intuition about quantum phenomena and focuses on helping them integrate qualitative and quantitative understanding. We also discuss findings from a preliminary in-class evaluation.

## INTRODUCTION

Quantum mechanics can be a challenging subject for students partly because it is unintuitive and abstract [1-6]. An experiment which has been conducted in undergraduate laboratories to illustrate fundamental principles of quantum mechanics involves the Mach-Zehnder Interferometer (MZI) with single photons [7]. We are developing a quantum interactive learning tutorial (QuILT) using *gedanken* (thought) experiments and simulations involving a MZI with single photons. The QuILT focuses on helping students learn topics such as the wave-particle duality of a single photon, interference of a single photon with itself, probabilistic nature of quantum measurements, and collapse of a quantum state upon measurement. Students also learn how photo-detectors (detectors) and optical elements such as beam-splitters in the path of the MZI with single photons affect the measurement outcomes.

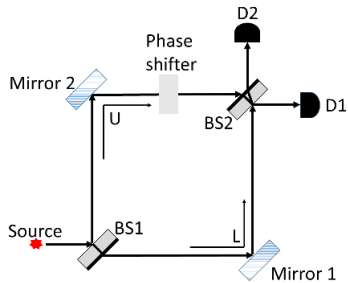

**FIGURE 1**. MZI setup with a phase shifter in the U path

Students are given a schematic diagram of the MZI setup in the QuILT (see the basic setup in Fig. 1). As students work through the QuILT, they are told to make simplifying assumptions about the MZI setup including the following: 1) all optical elements are ideal; 2) the non-polarizing beam-splitters (BS1 and BS2) are infinitesimally thin such that there is no phase shift when a single photon propagates through them; 3) the monochromatic +45º polarized single photons from the source travel the same distance in vacuum in the upper path (U) and lower path (L) of the MZI; and 4) the initial MZI without the phase shifter is set up such that there is a completely constructive interference at detector 1 (D1) because the photon state from both the U and L paths undergoes a phase change of $2\pi$ and arrives in phase at D1. There is destructive interference at detector 2 (D2) because the photon state undergoes a phase change of $2\pi$ in the U path and a phase change of $\pi$ in the L path and arrives out of phase at D2.

Using a guided approach to learning, the QuILT helps students reason about how observing interference of a single photon with itself at D1 and D2 can be interpreted in terms of not having "which-path" information (WPI) about the single photon [7]. WPI is a common terminology associated with these types of experiments popularized by Wheeler [8] and WPI is "known" about a photon if D1 and D2 can only project one component of the photon path state. For example, if BS2 is removed from the setup in Fig. 1, WPI is known for all single photons arriving at the detectors because only the component of a photon state along the U path can be projected in D1 and only the component of a photon state along the L path can be projected in D2. When WPI is known, each detector (D1 and D2) has equal probability of clicking. A detector clicks when a photon is detected by it and is absorbed (the state of the single photon collapses, i.e., the single photon state is no longer in a superposition of the U and L path states). When WPI is known, there is no way to know a priori which detector will click when a photon is sent until the photon state collapses either at D1 or at D2 with equal likelihood. On the other hand, WPI is unknown about single photons arriving at the detectors in the setup shown in Fig. 1 because BS2 mixes the path states of the single photon. Thus, D1 and D2 can project both components of the photon path state and the projection of both components at each detector leads to interference. When WPI is unknown and a large number of single photons are sent through the setup, if a phase shifter is inserted in one of the paths of the MZI (e.g., in the U path in Fig. 1) and its thickness is varied, the probability of photons arriving at D1 and D2 will change with the thickness of the phase shifter due to the interference of the components of the single photon state from the U and L paths. When WPI is known, changing the thickness of a phase shifter in one of the paths does not affect the probability of each detector clicking when photons are registered (equal probability for all thicknesses of phase shifter) [7].

## DEVELOPMENT OF THE QUILT

We developed a preliminary version of the QuILT (which includes a warmup with background information about the MZI setup with a beam of light and pre-/posttests to be given before and after the QuILT) that uses a guided approach to learning and accounts for the common student difficulties discussed later. The QuILT makes use of a computer simulation in which students can manipulate the MZI setup to predict and observe what happens at the detectors for different setups. Different versions of the QuILT were iterated with three physics faculty members several times to ensure that they agreed with the content and wording of the questions. We also administered it to several graduate students and upper-level undergraduate students to ensure that the guided approach was effective and the questions students must answer were unambiguously interpreted. Modifications were made based upon the feedback.

During the development of the QuILT, we investigated the difficulties students have with the relevant concepts including the wave-particle duality of a photon, interference of a single photon with itself, probabilistic nature of quantum measurements, and collapse of a quantum state upon measurement in order to effectively address them. We conducted 15 individual semi-structured think-aloud interviews with upper-level undergraduate and graduate students using different versions of an open-ended survey or earlier versions of the QuILT in which students were first asked to think aloud as they answered the questions related to the setup (including those with and without BS2) to the best of their ability without being disturbed. Later, we probed students further and asked them for clarification of points they had not made clear. Since both undergraduate and graduate students exhibited the same difficulties, we will not differentiate between the two groups further. Some common difficulties found in the interviews that were addressed in the QuILT included students struggling with the interference of a classical beam of light through the MZI, ignoring the wave nature of single photons, claiming that a photon is split into two photons after BS1 (see Fig.1), and struggling with how BS2 affects measurement outcomes.

**Difficulty with the interference of a beam of light at the detectors after passing through the MZI:** Interviews suggest that many students did not take into account the interference phenomenon of a classical beam of light. For example, regarding a beam of light with intensity $I$ propagating through the setup shown in Fig. 1, one student said: "There will be billions of photons in one beam so…approximately half go through U and half go through L. When going through BS2 they also have equal chance to reach D1 and D2. So the [intensity] on each [detector] will be $I/2$." Further probing indicates that students with these types of answers had some idea that a beam of light can be treated as a stream of photons but they often failed to invoke the wave nature of light which would lead, e.g., to constructive interference at D1 and destructive interference at D2 for the setup without phase shifter.

**Difficulties due to students' model that a single photon must act as a point particle:** Students struggled with the concept of wave/particle duality of a single photon and the fact that interference can be observed at the detectors due to a single photon state from the two paths (e.g., in Fig. 1, the photon state is in a superposition of the U and L path states after BS1 which can interfere at the detectors D1 and D2). Students often treated a single photon as a point particle, ignoring the wave-like nature of a single photon through the MZI. Some students claimed that a single photon can be split into two photons and it is these two photons that interfere at the detectors (instead of the fact that interference is due to the wave nature of single photons). For example, one student said "it seems like each photon with half of the energy of the incoming photon traveling along the U and L paths of the MZI is the only way for a photon to interfere with itself and have some probability of going through either path until getting measured." Other students claimed that neither the photon nor its energy will be split in half after BS1, but that each photon is localized in either the U or L path. These types of responses indicate that students struggled with the fact that a single photon can behave as a wave passing through the MZI and be in a superposition of U and L path states until a measurement is performed, e.g., at the detectors, and the single photon state collapses.

**Difficulty with the role of BS2:** Several students incorrectly claimed that either removing or inserting BS2 will not change the probability of the single photons arriving at each detector. For example, one student supplemented his claim as follows: "I don't see how BS2 affects/causes any asymmetry to make probabilities D1≠D2 or how BS2 causes a loss of photons." Another student who made similar incorrect claims about what happens at the detectors with and without BS2 said: "I say still 50% probability each since it's symmetric." Students who treated a single photon as a point particle and ignored its wave nature did not take into account the phase shifts affecting the component of the photon state from the U and L paths due to BS1 and BS2 (e.g., in Fig. 1) which influence the interference of the single photons at the detectors D1 and D2.

**Difficulty with how a detector collapses the single photon state**: Students often asserted that inserting an additional detector in either the U or L path of the MZI (not shown in Fig. 1) would not affect the interference at the detectors D1 and D2 at the end. They had difficulty with the fact that an additional detector, e.g., in the L path of the MZI in Figure 1 would collapse the state of the photon to the U or L path state so that D1 or D2 click with equal probability and the interference is destroyed. Instead, many students claimed that the photon state would remain delocalized in a superposition of the U and L path states (as in Fig. 1) and interference would be observed at D1 and D2. Some students correctly stated that a detector placed in the L path would absorb some photons but incorrectly claimed that there would still be interference displayed by the photons that reach D1 and/or D2. For example, one student said "Now path L is blocked [by a detector in the L path], so only ½ as many photons should hit the [detector D1 or D2 at the end]. I don't see how there can be any but constructive interference since path lengths are the same." Further probing of students with these types of responses suggests that they struggled with how placing a detector in the L path amounts to a measurement of path and destroys the delocalized single photon state which was in a superposition of the U and L path states before it reached the detector.

## PRELIMINARY EVALUATION

Once we determined that the QuILT was effective in individual administration, it was administered to 18 upper-level undergraduate students in a first-semester quantum mechanics course. Students were first given the pretest. They then worked through the QuILT in class and were asked to complete whatever they could not finish in class as homework. Then, they were given a posttest, which had the same questions as the pretest. Table 1 shows the common difficulties and percentages of students displaying them on the pre-/posttest questions and Table 2 displays the average percentage scores on pretest and posttest questions. The average normalized gain from pretest to posttest was 0.63 [9].

Question 1 on the pre-/posttest assessed student understanding of the classical interference of light in a situation in which a beam of light (instead of single photons) is sent through the MZI. Students were asked to explain why they agreed or disagreed with the following statement for the basic MZI setup without the phase shifter (Fig. 1): "If the source produces light with intensity $I$, the intensity of light at each point detector D1 and D2 will be $I/2$ each." The statement is incorrect because the MZI setup is such that there is completely constructive interference at D1 (the light from the U and L paths arrives completely in phase there with intensity $I$) and destructive interference at D2 (the light from the U and L paths is out of phase and no light arrives there). However, 78% of the students incorrectly agreed with this statement on the pretest (see Table 1), indicating that they did not take into account the interference phenomenon taking place at the detectors. After working on the QuILT, this difficulty was reduced.

**TABLE 1**. Common difficulties and percentages of students displaying them on the pre-/posttest questions.

| Q1 Ignoring interference phenomenon | 78/33 |
|---|---|
| Q2 BS1 causes the photon to split into two parts and halves the photon energy | 39/17 |
| Q2 Photon must take either U or L path | 22/6 |
| Q3 and Q4 Removing or inserting BS2 does not affect the probability of the detectors D1 and D2 registering photons | 56/17 |
| Q5 A photo-detector placed in the U or L path may absorb photons but does not affect whether interference is observed if photons arrive at detectors D1 and D2 | 28/0 |

**TABLE 2**. Average percentage scores on the pretest and posttest questions.

|  | Q1 | Q2 | Q3 | Q4 | Q5 |
|---|---|---|---|---|---|
| Pretest | 8 | 22 | 17 | 0 | 56 |
| Posttest | 58 | 81 | 72 | 44 | 100 |

Question 2 on the pre-/posttest assessed students' understanding of the wave nature of a photon. Students were asked to consider the following conversation between two students and explain why they agreed or disagreed with the statements: Student 1: "BS1 causes the photon to split in two parts and the energy of the incoming photon is also split in half. Each photon with half the energy travels along the U and L paths of the MZI and produces interference at the detectors." Student 2: "If we send one photon at a time through the MZI, there is no way to observe interference at the detectors. Interference is due to the superposition of waves from the U and L paths. A single photon must choose either the U or L path." Neither student is correct because a photon does not split into two parts with half the energy of the incoming photon but a single photon can be in a superposition of the U and L path states. 39% of the students agreed with Student 1 on the pretest. After working on the QuILT, this difficulty involving the splitting of photons was reduced (see Table 1). Furthermore, 22% of the students agreed with Student 2 in Question 2 on the pretest claiming that a photon must take either the U or L path. In the posttest, students performed better (see Table 1).

Questions 3 and 4 on the pre-/posttests assessed student understanding of the role of BS2. If BS2 is present, it mixes the state of the photon such that both U and L path components of the photon state can be projected at each detector and the photon interferes with itself at the detectors. In the setup students were given in which BS2 is present but the phase shifter is removed in Fig. 1, constructive interference occurs at D1 (the single photons always arrive at D1) and destructive interference occurs at D2 (no photon reaches D2). If BS2 is not present, the photon is still in a superposition of U and L path states after BS1 but each detector D1 and D2 can only project either the U or L path component of the photon superposition state. Thus, the photons do not display interference and each detector registers the photons with 50% probability. On the pretest, 56% of students incorrectly claimed that removing or inserting BS2 will not change the probability of the photon arriving at D1 and D2. This high percentage is consistent with the fact that these students did not acknowledge the wave nature and interference effects of single photons in response to other questions as well. Students often explicitly claimed that the photon behaves as a point particle, and each detector would register the photon with equal likelihood regardless of whether BS2 was present or not. In the QuILT, students learned that if BS2 is present, it evolves the state of the photon such that the photon state from both paths can be projected by each detector and interference is displayed at D1 and D2. On the posttest, students performed better (Table 1).

In Question 5 on the pre-/posttests, students were given a MZI with an additional detector placed in the L path between BS1 and BS2. They were then asked to describe how this situation compares to the situation in Fig. 1 in which no detector is present in the L path. In the new situation, if the additional detector in the L path does not absorb the photon, the photon path state must collapse to the U path. WPI is known and interference is not displayed by photons at D1 and D2. On the pretest, 28% of the students incorrectly claimed that adding a detector in the L path would not change anything or would only cause fewer photons arrive at detectors D1 and D2 because some photons are absorbed. These students struggled with the fact that the detector in the L path acts as a measurement device and will collapse the state of the photons not absorbed by it to the U path state. After working on the QuILT, the difficulty with the effect of an additional detector placed in the L path of the MZI was eliminated (see Table 1).

As shown in Table 2, many students still had difficulty with Questions 1 and 4 on the posttest. Question 1 relates to the interference phenomenon in the context of a beam of light that students were supposed to have learned about in the QuILT warmup at home (ungraded) before working on the QuILT about the MZI with single photons in class. In the future, the warmup should be administered as a graded homework to ensure that students complete it before working on the QuILT in class. Regarding the difficulty on Question 4 focusing on the role of BS2 on measurement outcomes, students who had difficulty on the posttest were often partially correct. Many correctly claimed that inserting BS2 would remove WPI but incorrectly claimed that the probabilities of detection of the photons at D1 and D2 would not change. For example, one student stated "the probabilities do not change, but we no longer have 'which-path' information about each incident photon." Some students displayed another difficulty and incorrectly claimed that D1 would register a photon 50% of the time and D2 would never register a photon because although the photon arrives at D2, destructive interference "kills" the photon. We have taken into account these findings from in-class administration in the next version of the QuILT. We are also developing an additional QuILT which strives to help students connect conceptual aspects of the MZI with single photons with mathematical formalism using a simple two state system involving photon path states.

## SUMMARY

The MZI QuILT focuses on helping students comprehend fundamental issues in quantum mechanics including the wave-particle duality of a single photon, interference of a single photon with itself, and how measurement collapses the delocalized superposition state of a single photon. In fact, many students in the class whose performance was discussed in the preceding section stated that it was one of their favorite QuILTs and they were excited to be introduced to contemporary topics in quantum mechanics. The preliminary evaluations are encouraging.

## ACKNOWLEDGEMENTS

We thank the National Science Foundation for awards PHY-0968891 and PHY-1202909.